\documentclass{ws-ijmpe}

\begin{document}

\catchline{}{}{}{}{}

\title{Mapping out the jet correlation
landscape: a perspective from PHENIX experiment}

\author{\footnotesize Jiangyong Jia for the PHENIX Collaboration}

\address{Chemistry Department, State University of New York at Stony Brook,\\
Stony Brook, NY 11794-3400, USA}
\address{Physics Department, Brookhaven National Laboratory,\\
Upton, NY 11796, USA\\
jjia@bnl.gov} \maketitle
\begin{history}
\end{history}

\begin{abstract}
This is a status report on where PHENIX stands in terms of mapping
out the landscape of jet correlation in $p_T$, hadron species,
$\sqrt{s}$. We discuss separately high $p_T$ correlation results
and low and intermediate $p_T$ correlation results. The former is
sensitive to the quenching of the jet by medium, the later allows
detailed study of the medium response to the (di-)jet.
\end{abstract}

\tableofcontents \markboth{Jiangyong Jia}{Jet correlation results
from PHENIX}

\section{Introduction}
One of the primary focus of the RHIC program is to study the
properties of the strongly interacting quark-gluon plasma (sQGP)
with high transverse momentum ($p_T$) jet and dijets. The
properties of the plasma can be deduced in two ways: on the one
hand, the suppression of single jets and di-jets signals at high
$p_T$ tells us about the stopping power or the transport
properties of the medium. On the other hand, study the the
response of the medium to the jets, especially their energy
dissipation in the medium, can reveal the collective properties or
equilibration process of the medium.

\begin{figure}[t]
\epsfig{file=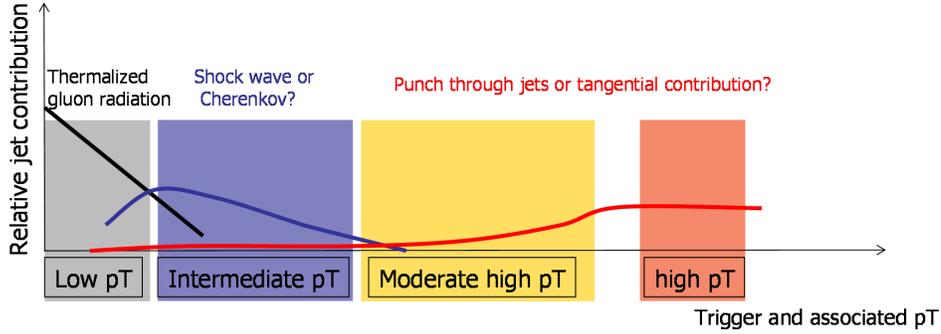,width=1.0\linewidth}
\caption{\label{fig:pic} A sketch of possible contributions to the
away side jet yield as function of $p_T$ of triggers and/or
partners. The $p_T$ axis is schematically divided into four
different regions. It is chosen such that the $p_{T,T}\geq
p_{T,A}$ (see Fig.\ref{fig:scan})).}
\end{figure}

The studies of two particle azimuth correlation at RHIC have
revealed detailed information on the jet interaction with the
medium, beyond what we learned from the single particle spectra
measurements. Experimentally, any deviations of the correlation
pattern in heavy-ion collisions from that in baseline p+p
collisions are attributed to some sort of medium effects. Such
deviations are found to be dependent on the $p_T$ of the trigger
hadrons ($p_{T,T}$) and partner hadrons ($p_{T,A}$).
Fig.\ref{fig:pic} show a schematic sketch of the landscape of the
two particle correlations in $p_{T,T}$ or $p_{T,A}$. Based on
existing experimental results and theoretical calculations, the
while $p_T$ range is divided into four regions, each of which is
probably sensitive to different physics. At high $p_T$,
STAR~\cite{Adams:2006yt} shows that the away side jet magnitude is
suppressed by factor of 4-5 relative to p+p, but maintaining a
relatively unmodified shape. It is consistent with the
fragmentation of the primary jets that survives the medium either
through tangential emission~\cite{Loizides:2006cs} or
punching-through mechanisms~\cite{Renk:2006nd} or
both~\cite{Zhang:2007ja}. At intermediate $p_T$ range(1-4
GeV/$c$), the away side jet is broadened~\cite{Adler:2005ee} or
even concave like~\cite{Adare:2006nr} in shape but its amplitude
is enhanced relative to p+p~\cite{Adams:2005ph}. The features were
believe to be the consequence of the response of the medium to the
energy degradation of the high $p_T$ jets. Many theoretical
scenarios have been proposed in the last few years to describe the
energy dissipation processes, including gluon
radiation~\cite{Vitev:2005yg}, Cherenkov
radiation~\cite{Dremin:1979yg,Koch:2005sx}, ``Mach Shock''
mechanism~\cite{Casalderrey-Solana:2004qm}. But deflected jet
pictures such as hydrodynamical wake~\cite{Armesto:2004vz} or
multiple scattering of the jet by the medium~\cite{Chiu:2006pu}
were proposed as well. The first two models require additional
particle production, whereas the other models require
redistribution of the momentum/energy among existing particles.

In between these two $p_T$ regions, experimental data show an
suppressed and possibly very flat away side with no visible peak
structure~\cite{Adler:2002tq}. This region (we call moderately
high $p_T$ region) could have contributions from both jet
fragmentation and responses of the medium. The flat but still
positive structure could be the combined result of a fragmentation
contribution concentrated around $\Delta\phi=\pi$ and a medium
contribution centered at 1 radian away from $\pi$. The flat
structure probably implies the two components have roughly equal
yield at this $p_T$ region.

%

Lastly is the low $p_T$ region. In most jet quenching
models~\cite{Salgado:2003rv,Vitev:2005yg}, the bremsstrahlung
radiation was believed to be largely contained in a narrow angular
range (0.5 rad) around the jet axis. These gluons might be the
precursor of the collective medium response, but it is also
conceivable that some gluons simply thermalized with the medium
after additional scattering. In this picture, one would also
expect an enhancement in the region around $\pi$ at low $p_T$ and
it should have an exponential thermal shape.

Given that the correlation patterns depend strongly on the $p_T$
of both particles in the pairs, it is crucial to map out the
awayside jet properties in a broad $p_T$ range and to identify the
transition regions between different correlation patterns. We
focus separately on the high $p_T$ region, which is ideal for
studying jet quenching and jet tomography, and on intermediate
$p_T$ region, which is dominated by the medium response. PHENIX
results on hadron-hadron, identified hadron-hadron and three
particle correlation are discussed in this context, with an
emphasize on new results we have shown in QM2006.

\section{$p_T$ evolution of correlation pattern}
In correlation analysis, one correlate hadrons in one $p_T$ region
(``trigger'') with those in another $p_T$ window (``partner'').
The hadron pairs from same jet tend to appear at
$\Delta\phi=|\phi_A-\phi_B|\sim0$ and those from for back-to-back
dijet tend to appear at $\Delta\phi\sim\pi$. We are interested in
the partner yield distribution per-trigger, $Y_{\rm{jet}}$ (PTY),
which is defined in a way similar to the previous $d+Au$ and $p+p$
analysis\cite{Jia:2004sw,Adler:2005ad}:
\begin{eqnarray}
\label{eq:core1}\nonumber Y_{\rm{jet}} = \frac{\int d\Delta\phi
N^{\rm{mix}}}{2\pi
N_{t}\epsilon}\left(\frac{N^{\rm{fg}}(\Delta\phi)}{N^{\rm{mix}}(\Delta\phi)}-
\xi\left(1+2v_2^{t} v_2^{a} \cos2\Delta \phi\right)\right)
\end{eqnarray}
with the additional $2v_2^{t} v_2^{a} \cos2\Delta \phi$ term to
take into account the flow modulation of the background pairs in
Au+Au collisions. $N_{t}$ is the number of trigger, $\epsilon$ is
the efficiency for associated hadrons in full azimuth and in
$\pm0.35$ pseudo-rapidity. $N^{\rm{fg}}(\Delta\phi)$ and
$N^{\rm{mix}}(\Delta\phi)$ are same-event pair and mixed-event
pair distributions, respectively. The superscript $t$ and $a$
stands for the trigger and partner particles. $\xi$ is a
normalization factor which is the ratio of the combinatorics pairs
in the same event to those in the mixed event. $\xi$ is typically
bigger but very close to 1.

\begin{figure}[h]
\epsfig{file=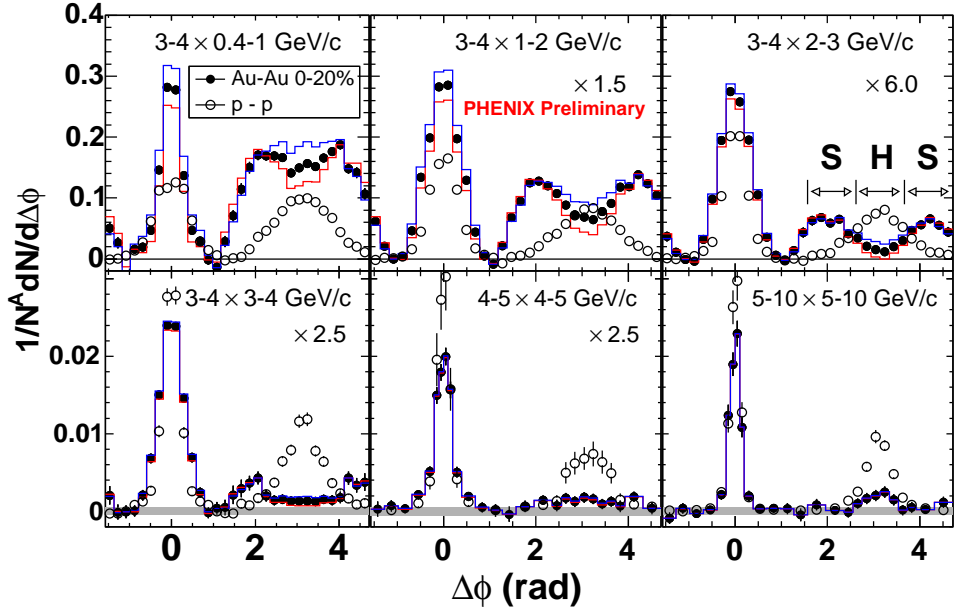,width=1.0\columnwidth}
\caption{\label{fig:shape} The yield of associated hadron
per-trigger in $\Delta\phi$ for successively increasing $p_{T,T}$
and $p_{T,A}$ in 0-20 \% Au+Au collisions. The two lines around
the data indicate the errors due to uncertainty of elliptic flow;
the shaded bands around 0 indicate the uncertainty from ZYAM
method.}
\end{figure}

Fig.\ref{fig:shape} shows the PTY for various combination of
trigger and partner $p_T$ in central Au+Au collisions, in
comparison with p+p. The partner $p_T$ is constrained to be less
than trigger $p_T$. From left to right and top to bottom, the
$p_T$ ranges of one or both particles increase, going from
$3-4\times0.4-1$ GeV/$c$ (trigger $p_T$ vs partner $p_T$) to
$5-10\times5-10$ GeV/$c$. The solid lines around the points
represent the error due to the elliptic flow uncertainty, which
peaks at 0 and $\pi$ and is the dominating error at $p_T<3$
GeV/$c$. The shaded bands indicate the error due to uncertainty on
the background level which is fixed by the Zero Yield At Minimum
(ZYAM) method~\cite{Ajitanand:2005jj}. This error depends on the
statistical uncertainty at the ZYAM minimum and is the dominating
at $p_T>3$ GeV/$c$.

The top row shows the jet yield in three successive partner $p_T$
ranges covering 0.4 to 3 GeV/$c$ with trigger $p_T$ fixed in 3-4
GeV/$c$. The away side jet shapes are dramatic different between
Au+Au and p+p. The p+p data always peak at $\pi$; The Au+Au data
show a concave shape with a dip around $\pi$ and two side peaks
around 1 radian from $\pi$. The concaveness grows with increasing
partner $p_T$, however the location of the side peaks is not
changing. To characterize the distorted away side jet shape, we
divided the away side into a ``Head'' region
($\pi/6<|\Delta\phi-\pi|$) and a ``Shoulder'' region
($\pi/6<|\Delta\phi-\pi|<\pi/2$), as illustrated in Fig.1c. They
are so named because for normal jet as in p+p, the ``Head'' region
contains bulk of the jet fragmentation, whereas the ``Shoulder''
contains the tail of the jet fragmentation and the radiative
contribution. The ``Head'' region is sensitive to the level of jet
quenching while the ''Shoulder'' region is suitable for studying
the medium response. Clearly, the behavior of the Au+Au data in
the two regions are dramatically different from that of p+p.
Namely there a suppression in the ``Head'' region and an
enhancement in the ``Shoulder'' region. The suppression in the
``Head'' region sets in around 1 GeV/$c$ and grows with larger
partner $p_T$, while the enhancement in the ``Shoulder'' region
persists up to 4 GeV/$c$.

In the bottom panels of Fig.\ref{fig:shape}, as the $p_T$ of both
hadrons are further increased, the awayside ``displaced'' peaks in
Au+Au data seem to be compressed relative to p+p and the Au+Au
near side. This reflects the fact that the away side yield drops
faster than the near side with increasing partner $p_T$. In 4-5
GeV/$c$ bin, the Au+Au away side is broader than p+p but is no
concave. At the highest $p_T$ bin (5-10 GeV/$c$), the away side
turns into a convex shape, similar in shape to p+p but it's
magnitude is largely suppressed.

Note that medium modification is on the pairs. Thus the shape of
the away side should be symmetric with respect to the trigger
$p_T$ and partner $p_T$. Fig.\ref{fig:shape} suggests that the
away side can be categorized in following four regions depends on
the $p_{T,T}$ and $p_{T,A}$ based on its shape: 1) a flat region
at $p_{T,T}$ or $p_{T,A}<1$ GeV/$c$; 2) a concave shape region at
$1\lesssim p_{T,T},p_{T,A}\lesssim 4$ GeV/$c$; 3) a convex region
$p_{T,T},p_{T,A} \gtrsim 5 $ GeV/$c$; 4) a transition region
(almost flat away side) $p_{T,T}(p_{T,A})\gtrsim 5 $ GeV/$c$
$p_{T,A} (p_{T,T})\lesssim 3 $ GeV/$c$.

\section{Medium response}
\subsection{Hadron-hadron correlation}
\begin{figure}[ht]
\epsfig{file=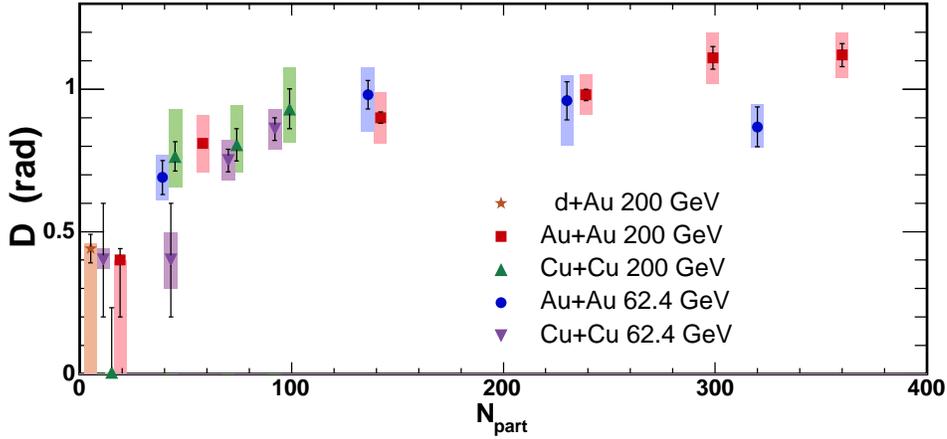,width=1.0\columnwidth}
\caption{\label{fig:d} Collision centrality, energy, and system
size dependence of shape parameters ``D''.}
\end{figure}

PHENIX have tried different methods to quantify the behavior of
the away side concave shape at intermediate $p_T$. The location of
the side peaks are characterized by ``D''~\cite{Adare:2006nr}: the
distance of the peak to $\pi$. ``D'' is found by triple gauss
function fit of the away side:
\begin{eqnarray}
J(\Delta\phi) =
G_N(\Delta\phi)+G_A(\Delta\phi-\pi-D)+G_A(\Delta\phi-\pi+D)
\end{eqnarray}
Fig.\ref{fig:d} show D for several collision systems and
collisions energies. The position of the peaks was found to be
around 1 radian from $\pi$ independent of the centrality (
$N_{\rm{part}}>$100), collision energy and system size. In
addition, ``D'' was found to be relatively insensitive to the
change of partner $p_T$~\cite{Adare:2006nr}, which is also
confirmed in Fig.\ref{fig:shape}. This suggests that ``D''
reflects some kind of universal properties of the medium, such as
the speed of sound as suggested by the Mach Cone mechanism. It
can't be described by Cherenkov gluon radiation
models~\cite{Dremin:1979yg,Koch:2005sx} or most of the deflected
jet models such as hydrodynamical wake~\cite{Armesto:2004vz} or
Markov random scattering~\cite{Chiu:2006pu}. Such mechanisms would
predict narrowing of the peak angle with increasing $p_T$.

PHENIX also try to characterize the convexity of the away side jet
shape with the ratio of average jet yield in the ``Head'' region
to that in the ``Shoulder'' region, $R_{\rm{HS}}$,
\begin{eqnarray}
R_{\rm{HS}}  = {{\frac{{\int_{\rm{Head}} {d\Delta \phi
Y_{\rm{jet}} (\Delta \phi )} }} {\rm{Head}}} \mathord{\left/
 {\vphantom {{\frac{{\int_{\rm{Head}} {d\Delta \phi Y_{\rm{jet}} (\Delta \phi )} }}
{\rm{Head}}} {\frac{{\int_{\rm{Shoulder}} {d\Delta \phi
Y_{\rm{jet}} (\Delta \phi )} }} {\rm{Shoulder}}}}} \right.
 \kern-\nulldelimiterspace} {\frac{{\int_{\rm{Shoulder}} {d\Delta \phi Y_{\rm{jet}} (\Delta \phi )} }}
{\rm{Shoulder}}}}
\end{eqnarray}
Each of the distribution in Fig.\ref{fig:shape} produces one value
of $R_{\rm{HS}}$ that captures essence the the away side jet
shape. In general, one expect $R_{\rm{HS}}<1$ for a concave shape,
$R_{\rm{HS}}\approx 1$ for a flat distribution and $R_{\rm{HS}}>1$
for a convex shape. Thus $R_{HS}$ is an ideal quantity for
studying the transition between the ``Shoulder'' dominated region
to ``Head'' dominated region in $p_T$. Note $R_{\rm{HS}}$ is
purely a shape variable, thus it is symmetric with respect to
$p_{T,T}$ and $p_{T,A}$, i.e.
$R_{\rm{HS}}(p_{T,T},p_{T,A})=R_{\rm{HS}}(p_{T,A},p_{T,T})$.

\begin{figure}[ht]
\epsfig{file=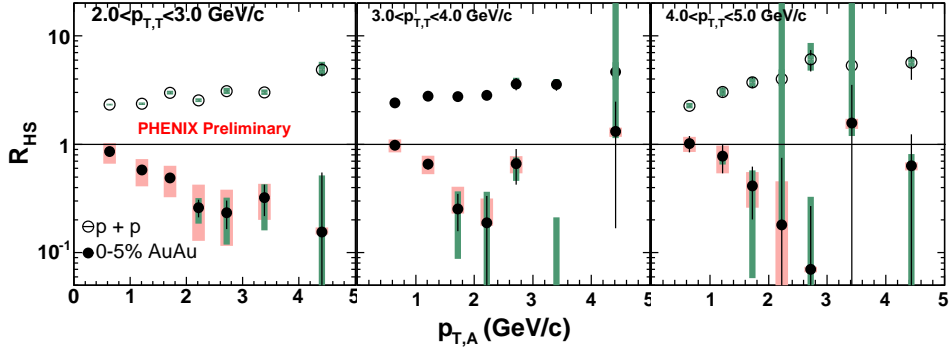,width=1.0\columnwidth}
\caption{\label{fig:ratio} $R_{\rm{HS}}$ (The ratio of the average
yield in head and shoulder region) as function of $p_{T,A}$ in
three $p_{T,T}$ ranges for 200 GeV 0-5\% central Au+Au and p+p.
Concave $R_{\rm{HS}}<1$, flat $R_{\rm{HS}}\approx1$, convex
$R_{\rm{HS}}>1$.}
\end{figure}

Fig.\ref{fig:ratio} summarize the $p_{T,T}$ and $p_{T,A}$
dependence of $R_{\rm{HS}}$ in p+p and central Au+Au collisions.
It is presented in fine bins of partner $p_{T}$ for three trigger
$p_{T}$ regions. The wide red shaded error bars represent the
elliptic flow error, which is correlated in $p_T$; The narrow
green shaded bar represent the ZYAM error, which can vary from
point to point.
In p+p collisions, $R_{\rm{HS}}$ is always above one and increases
with $p_{T,A}$. This suggests that the away side is always peaked
around $\pi$ in p+p and its width narrows with increasing
$p_{T,A}$.
The same ratio in 0-5\% central Au+Au collisions are drastically
different. For trigger $p_T$ in 2-3 GeV/$c$, $R_{\rm{HS}}$
decrease from around 1 at $p_{T,A}<1$ GeV/$c$ to a level of
$0.3\pm0.1$ at 3-4 GeV/$c$. For trigger $p_T$ in 4-5 GeV/$c$, the
decrease of $R_{\rm{HS}}$ is less significant. The level at low
partner $p_T$ is around 1, relatively insensitive to the trigger
$p_T$. This suggests that away side is flat at low partner $p_T$
independent of the trigger $p_T$.

\begin{figure}[th]
\epsfig{file=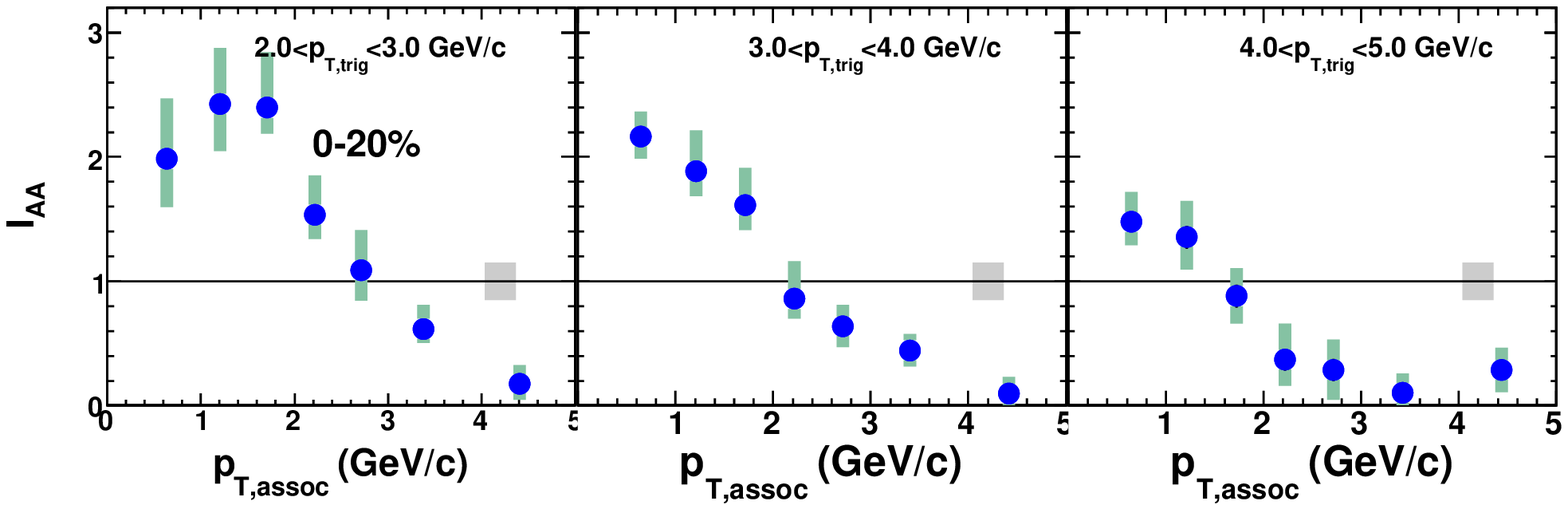,width=1.0\columnwidth}
\epsfig{file=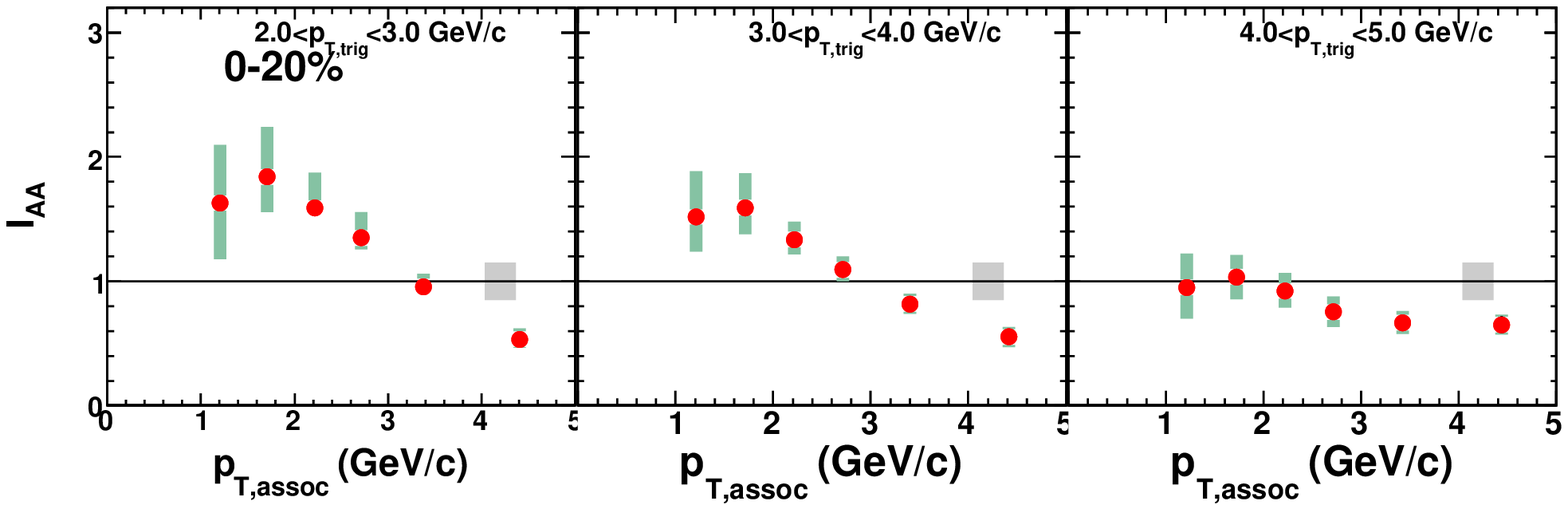,width=1.0\columnwidth}
\caption{\label{fig:inte} $I_{\rm{AA}}$ for three different
trigger bins in the away side (top panels) and the near side
(bottom panels) in 0-20\% central Au+Au collisions at 200 GeV.}
\end{figure}

Fig.\ref{fig:shape} suggests that not only the away side jet shape
but also the jet multiplicities are modified in central Au+Au
collisions relative to that in p+p. The modifications depend on
trigger $p_T$, partner $p_T$ and $\Delta\phi$ and can be
quantified by $I_{\rm{AA}}$, the ratio of PTY in Au+Au to that in
$p+p$ within a certain $\Delta\phi$ window W:
\begin{equation}
I_{\rm{AA}}^W  = {{\int_{\Delta\phi\in W} {d\Delta \phi
Y_{\rm{jet}}^{\rm{Au + Au}} } } \mathord{\left/
 {\vphantom {{\int_{\Delta\phi\in W} {d\Delta \phi Y_{\rm{jet} }^{\rm{Au + Au}} } } {\int_{\Delta\phi\in W} {d\Delta \phi Y_{\rm{jet} }^{p + p} } }}} \right.
 \kern-\nulldelimiterspace} {\int_{\Delta\phi\in W} {d\Delta \phi Y_{\rm{jet} }^{p + p} } }}
\end{equation}
The top panels Fig.\ref{fig:inte} show the $I_{\rm{AA}}$ as
function of partner $p_T$ at away side ($|\Delta\phi-\pi|<\pi/2$)
for three trigger $p_T$ bins. $I_{\rm{AA}}$ is greater than 1 at
low partner $p_T$ but drops towards high partner $p_T$. This
implies that central Au+Au data have a significant enhancement at
low partner $p_T$ and a strong suppression at high partner $p_T$.
For higher trigger $p_T$, the enhancement at low partner $p_T$ is
smaller and the suppression at high partner $p_T$ is stronger. The
point where $I_{\rm{AA}}$ cross 1 shifts to lower partner $p_T$,
indicating a stronger decrease of $I_{\rm{AA}}$ in partner $p_T$
for higher trigger $p_T$. The observed $p_{T,T}$ and $p_{T,A}$
dependence is a consequence of the competition between the
enhancement in the ``Shoulder'' region and the suppression in the
``Head'' region shown in Fig.\ref{fig:shape}. At low $p_{T,A}$,
the away side is (thus $I_{\rm{AA}}$) is dominated by the
``Shoulder'' region; at sufficiently high $p_{T,A}$, the away side
is dominated by the ``Head'' region. The point where
$I_{\rm{AA}}=1$ would depend on both trigger and partner $p_T$.
For 4-5 GeV/$c$ trigger, the suppression level seems to saturates
around 0.3 at $p_{T,A}>3$, close to the $R_{\rm{AA}}$ value for
4-5 GeV/$c$ inclusive hadrons.

\begin{figure}[th]
\epsfig{file=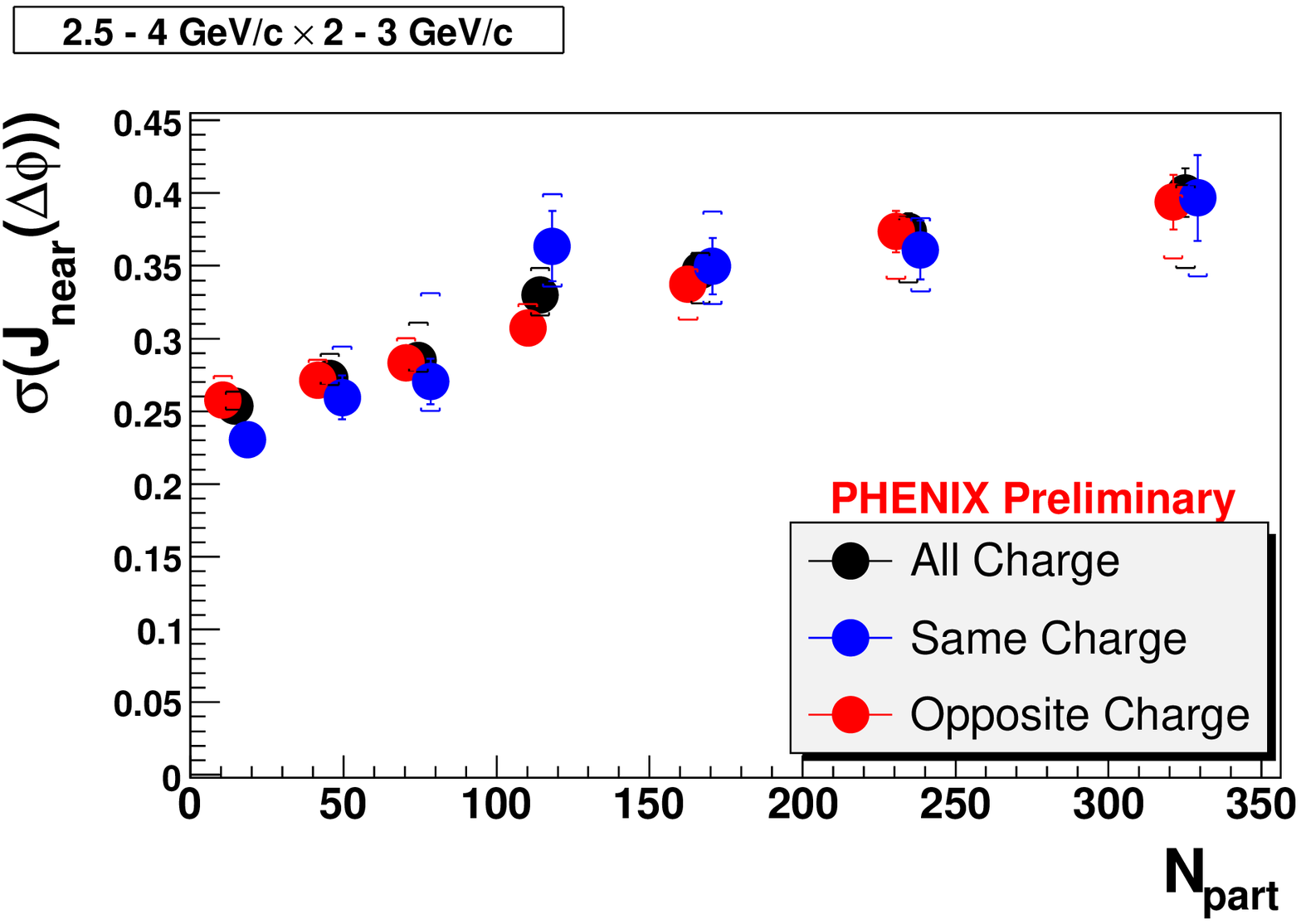,width=0.54\columnwidth}
\epsfig{file=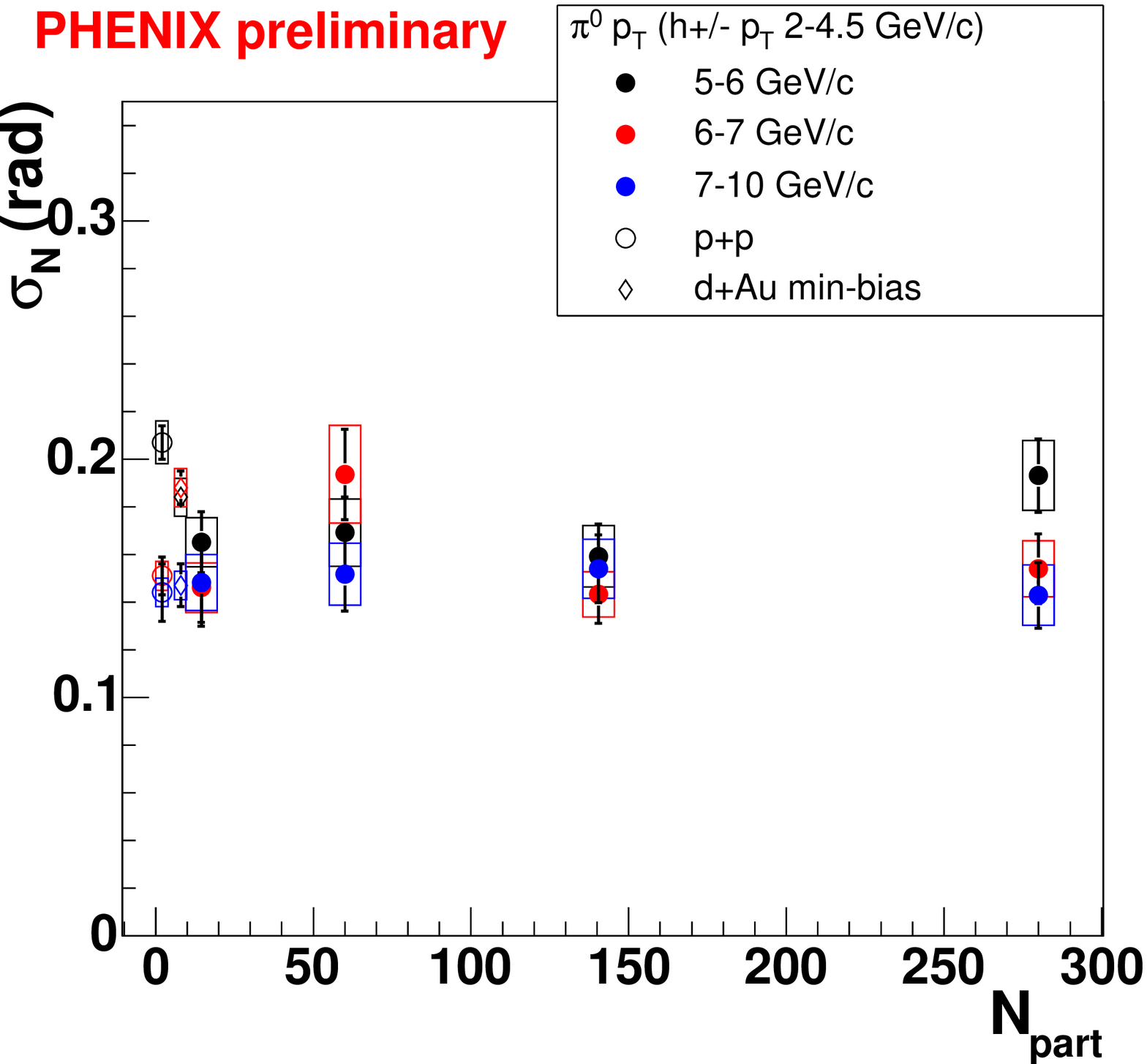,width=0.4\columnwidth}
\caption{\label{fig:width} Centrality dependence of the near side
gauss width in 200 GeV Au+Au collisions at intermediate $p_T$ of
$2.5-4\times2-3$ GeV/$c$ (Left) and at various high $p_T$ bins
above 5 GeV/$c$ (Right).}
\end{figure}

The bottom panels of Fig.\ref{fig:shape} show the $I_{\rm{AA}}$ as
function of partner $p_T$ at away side ($|\Delta\phi|<\pi/3$) for
the same three trigger $p_T$ bins. We see similar enhancement at
low partner $p_T$, and the enhancement diminishes towards high
partner $p_T$. As one increase the trigger $p_T$, the dependence
on partner $p_T$ weakens, and the $I_{\rm{AA}}$ distributions
become flatter in partner $p_T$. The near side enhancement has
been observed by STAR~\cite{Adams:2005ph} and PHENIX
~\cite{Adler:2004zd,Adler:2005ee}, this enhancement was attributed
by STAR to the a ridge contribution which extends out in
$\Delta\eta$ up to $\pm2$. The ridge contribution seen by STAR in
$|\Delta\eta|$<1.7 is about factor of three compare to the pure
jet contribution, and was argued to be reason for the near side
enhancement. The enhancement seen in PHENIX is smaller than
observed in STAR, because of the ridge yield in PHENIX's limited
$\eta$ acceptance is smaller. The decrease of the enhancement with
increasing trigger $p_T$ suggests that the ridge component is
softer than the jet component. The $p_T$ range where the ridge
yield is important seems to be in $p_{T,T},p_{T,A}<4$ GeV/$c$,
very similar to the range for the enhancement at the away side.
PHENIX have also observed a broadening of the near side jet width
in central Au+Au collisions~\cite{Jia:2005ab} at intermediate
$p_T$, the broadening was found to disappear at high trigger
$p_T$~\cite{Grau:2005sm} (also see Fig.\ref{fig:scan}). These
observations are summarized in Fig.\ref{fig:width}. The width
results are also suggestive of a ridge contribution that is
important at intermediate $p_T$ but disappears at large $p_T$. Due
to its limited $\eta$ acceptance, PHENIX so far haven't be able to
observe the ridge signature directly. However, the detailed study
of near side jet shape and yield in broad $p_T$ range can provides
indirect but still valuable constrains on the properties of the
ridge.

\begin{figure}[ht]
\epsfig{file=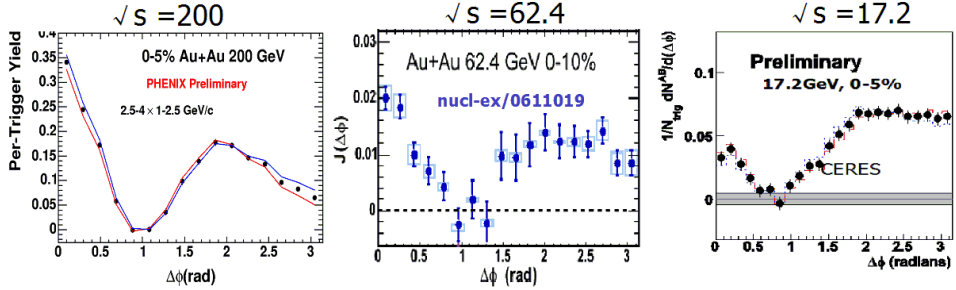,width=1.0\columnwidth}
\caption{\label{fig:jetsqrt} a) Per-trigger yield in central Au+Au
collisions at $\sqrt{s_{\rm{NN}}} =$ 200 GeV from PHENIX. b) The
extract jet function at $\sqrt{s_{\rm{NN}}} =$ 62.4 GeV from
PHENIX. c) Per-trigger yield at $\sqrt{s_{\rm{NN}}} =$ 17.3 GeV
from CERES.}
\end{figure}

PHENIX have carried out the intermediate $p_T$ correlation
analysis at both $\sqrt{s_{\rm{NN}}}=200$ and 62.4 GeV
~\cite{Adare:2006nr}. CERES also carried out similar analysis at
$\sqrt{s_{NN}}=17.2$ GeV selection~\cite{Ploskon:2007es}. The
results from the three collision energies for
$1<p_{T,A}<2.5<p_{T,T}<4$ GeV/$c$ are summarized in the
Fig.\ref{fig:jetsqrt}. The middle panel~\cite{Adare:2006nr} shows
the ``jet fraction'', so the scale can't be compared directly with
the other two panels which show the per-trigger yield. But the jet
shape can be directly compared among the three distributions.
Fig.\ref{fig:jetsqrt} indicate a clear modification of away side
jet at all energies. They all have displaced peaks around 1 radian
from $\pi$, the concaveness of the away side distributions seems
to decrease for lower $\sqrt{s}$. Since the trigger selection are
similar, we expect the away side jet energy would be mostly
determined by the trigger $p_T$, modulo a weak $\sqrt{s}$
dependence of the trigger bias (reflected by $\langle z\rangle$)
due to the $\sqrt{s}$ dependence of the parton spectra shape.
CERES Time Projection Chamber $\eta$ range is 2.1-2.7 in lab
frame, which corresponds to 0.1-0.7 in CM frame. Thus its rapidity
range is 0.6 and is close to PHENIX range of 0.7. Thus the jet
yield at 17.3 GeV can be compared directly with that at 200 GeV
after multiplied by 0.7/0.6 = 1.17 to correct for the difference
in $\eta$ acceptance. The maximum and minimum value is 0.17 and
0.07 for 0-5\% central Au+Au collisions at 200 GeV
(Fig.\ref{fig:jetsqrt}a) and 0.08 and 0.07 for 0-5\% Pb+Pb
collisions at 17.3 GeV after the acceptance correlation
(Fig.\ref{fig:jetsqrt}c). Thus the amplitude of the displaced peak
in CERES is about factor 2 lower than PHENIX value, whereas the
level at the ``Head'' region in CERES is surprisingly close to
PHENIX value. This probably suggests that the yield in the
``Head'' region is dominated by fragments from jet. The jet
multiplicity are somewhat higher at RHIC due to smaller $\langle
z\rangle$, but jet quenching is stronger at RHIC than that in SPS.
The yield at ``Shoulder'' region is significantly smaller at SPS
than at RHIC, possibly suggesting a weaker medium effect. Clearly
detailed study of the $\sqrt{s}$ dependence of the ``Head'' and
``Shoulder'' yield can provide crucial constrains on interplay
between jet quenching and medium response.

\begin{figure}[ht]
\epsfig{file=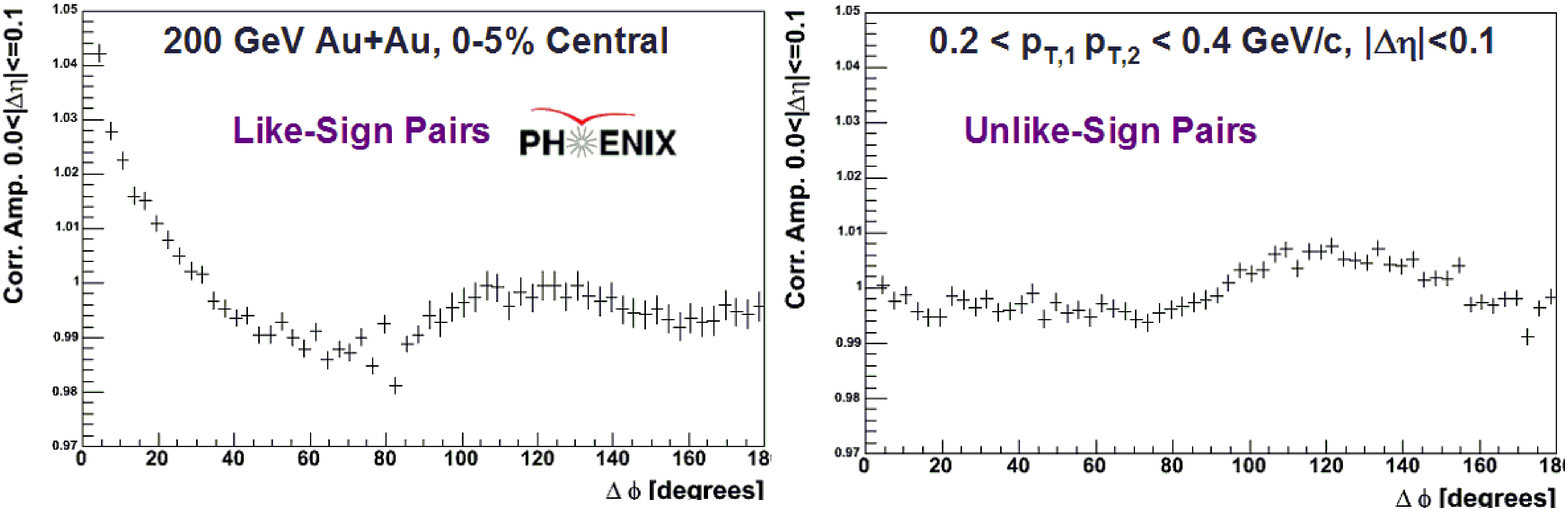,width=1.0\columnwidth}
\caption{\label{fig:lowpt1} Low $p_T$ like-sign (left) and
unlike-sign correlation function for 0-5\% central 200 GeV Au+Au
collisions.}
\end{figure}

PHENIX also obtained some interesting but puzzling results for
correlations between very low $p_T$ hadrons ($0.2<p_T<0.4$
GeV/$c$)~\cite{Mitchell:2007wz}. Fig.\ref{fig:lowpt1} shows
azimuthal correlation in 0-5\% Au+Au collisions for like-sign
pairs (left) and unlike-sign pairs (right). The like-sign
correlation shows a near side peak consistent with HBT
correlation, whereas unlike-sign correlation does not. What is
surprising is the observation of a displaced peak at a location
that is consistent with what has been observed for high $p_T$
correlation. Whether it has the same physics origin is not clear.
If this is due to correlation between soft particle in the jets or
mini-jets, why the unlike-sign correlation does not show any
enhancement at the near side as well (the ridge argument)? It is
also worth pointing out that the displaced peak mostly
concentrated in a very narrow window of $|\Delta\eta|<0.1$ and
that the peak only shows up in large $N_{\rm{part}}\gtrsim 100$
and is insensitive to the collision energy. It is observed at
0-5\% Au+Au at 62.4 GeV but not in central Cu+Cu collisions at 200
or 62.4 GeV. Further characterization of the shape and yield of
this away side peak is underway to clarify the picture.

\subsection{Three particle correlation}
\begin{figure}[h]
\epsfig{file=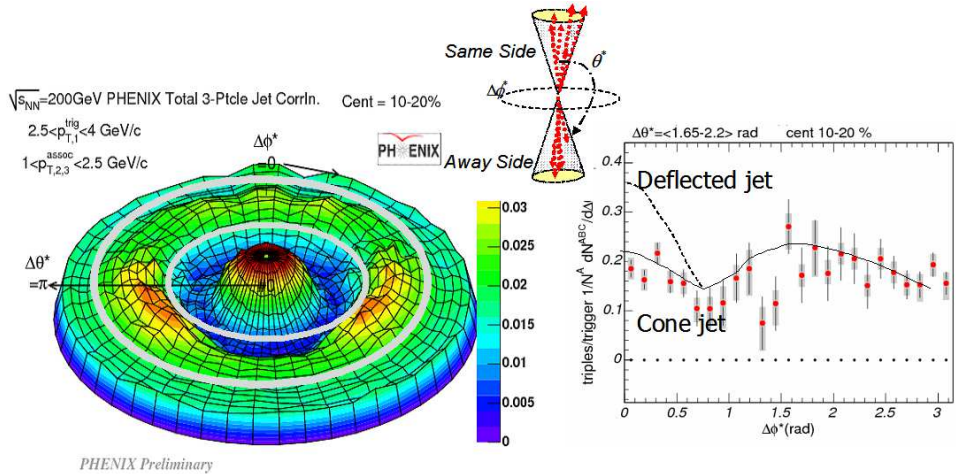,width=1.0\columnwidth}
\caption{\label{fig:3p} (Left) Full $\Delta\phi^*$,
$\Delta\theta^*$ correlation surface in 10-20\% Au+Au collisions.
(Middle insert) Schematic illustration of coordinate system.
(Right) the 1-D projection of along $\Delta\phi*$ for
$1.65<\Delta\theta^*<2.2$ rad.}
\end{figure}

Three-particle correlation can provide direct topological features
on the interaction of the jets with the medium, thus potentially
allows clear distinction of various modification patterns such as
Mach-cones and Deflected jets which can't be directly
differentiated with two particle correlation method. There are
four independent angular variables among the three particles,
$\Delta\phi_{12}= \phi_1-\phi_2, \Delta\phi_{13}=\phi_1-\phi_3$,
$\Delta\eta_{12}$ and $\Delta\eta_{13}$. If one is mainly
interested in the azimuthal correlation, as is done in
STAR~\cite{Ulery:2006ha} and in earlier PHENIX
measurements~\cite{Ajitanand:2005xa}, then there are only two
independent variable $\Delta\phi_{12}$ and $\Delta\phi_{13}$.
Recently PHENIX has chosen to characterize three particle
correlations through the use a local polar coordinate system where
the trigger direction is the z axis (see insert of
Fig.\ref{fig:3p})~\cite{Ajitanand:2006is}. In this scheme, the
four independent variables are $\theta_1^*$,$\theta_2^*$,
$\phi_1^*$ and $\phi_2^*$. Note that the coordinate system is
defined for each trigger so it changes trigger by trigger relative
to the global frame. In the data analysis, PHENIX chose
$\Delta\theta^*=\theta_1^*-\theta_2^*$ and
$\Delta\phi^*=\phi_1^*-\phi_2^*$ as the two independent variables.
These two variables combine the information of all four angles,
thus provide a very different way of presenting the jet signal.

Left panel of Fig.\ref{fig:3p} shows the resulting three-particle
correlation function in $\Delta\theta^*$ and $\Delta\phi^*$. The
soft-soft and soft-hard flow backgrounds have been subtracted, but
it still includes the 2+1 (two particles from di-jet and one from
background) flow background. It is important to note that this
correlation function as generated also contains the effect of the
PHENIX acceptance. One can see three peaks in the ring region
($1.65<\Delta\theta^*<2.2$ rad) between the two gray circles,
located at $\Delta\phi^*\approx 0$ and
$\Delta\phi^*\approx\pm\pi/2$. These peaks were not observed for
simulation where only normal di-jet is included, thus are
signatures for the medium effects. After removing the 2+1 flow
contributions, right panel of Fig.\ref{fig:3p} shows the
projection along the ring on the $\Delta\phi^*$ axis, together
with expectation of simplistic cone jet and deflected jet
scenarios given by Monte-Carlo simulations. In both scenarios, one
expected a narrow peak around $\Delta\phi^*\approx 0$ and a broad
peak around $\Delta\phi^*\approx\pi/2$. The difference is that the
deflected jet scenario predicts a bigger jet amplitude at
$\Delta\phi^*\approx 0$ than that for the cone jet. From right
panel of Fig.\ref{fig:3p}, it appears that the PHENIX measurements
are most consistent with the excitation of a Mach cone in the
medium and not consistent with a deflected jet hypothesis.

\subsection{Identified hadron-hadron correlation}

Previous results indicate that the jet properties at the nearside,
awayside ``Head'' and awayside ``Shoulder'' regions are quite
different. The near side is consist of a ridge component along
$\Delta\eta$ and a jet fragmentation contribution. The awayside
``Shoulder'' region is dominated by collective medium response and
the ``Head'' region seems to be consist of jet fragmentation plus
possible feedback from the radiated gluons. Given the complexity
of the problem and many theoretical possibilities, one need
additional handles to test this picture. Correlation with
identified particles can reveal the composition or the chemistry
of the jet yield in the three regions, thus provide valuable
constraints on the underlying physics.

\begin{figure}[th]
\begin{center}
\epsfig{file=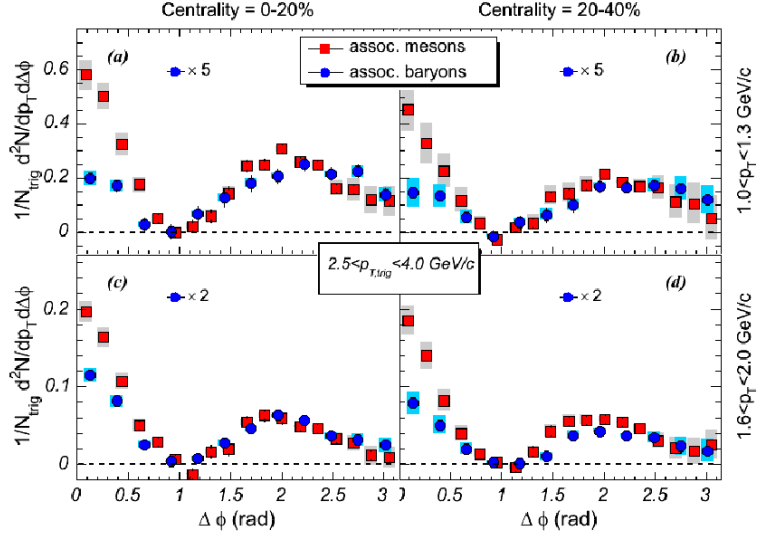,width=0.8\columnwidth}
\caption{\label{fig:id3} The associated meson and baryon
$\Delta\phi$ distribution for unidentified charged hadron
triggers.}
\end{center}
\end{figure}

Fig.\ref{fig:id3} shows the partner meson and baryon yield in
1-1.3 GeV/$c$ (top panels) and 1.6-2.0 GeV/$c$ (bottom panels) for
unidentified trigger in 2.5-4 GeV/$c$. Results for 0-20\% and
20-40\% Au+Au collisions are shown in the left panels and right
panels, respectively. The partner baryon yields are scaled to
approximately match the partner meson yields. Clearly, the away
side shows a concave shape for both partner mesons and baryons.
This is not surprising for partner meson since it dominates the
charged hadrons. But it is interesting to see that partner baryons
also have a similar but less concave shape (higher in the head
region and lower in the shoulder region in the figure).

\begin{figure}[th]
\begin{center}
\epsfig{file=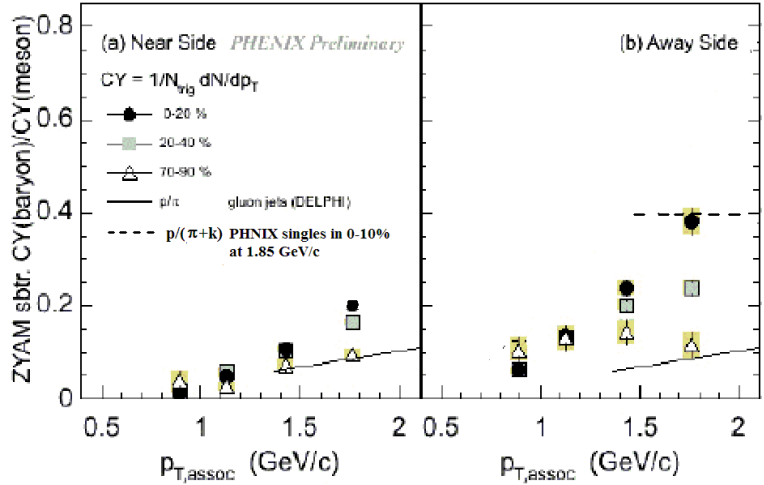,width=0.8\columnwidth}
\caption{\label{fig:id2} Ratio of partner baryon to meson at the
near side (left panel) and away side (right panel). It is plotted
as function of partner $p_T$ for 0-20\%, 20-40\% and 70-90\%
centrality bin. Trigger particles are charged hadrons with $2.5<
p_T <4.0$ GeV/$c$.}
\end{center}
\end{figure}

Fig.\ref{fig:id2} shows the ratios of the parter baryon/meson
yield as function of partner $p_T$ for several centrality bins at
the near side (left panel) and away side (right panel). The ratios
are compared with that for the inclusive spectra and $e^+e^-$
collisions. The ratio for the near and away side jet grow steadily
with increasing partner $p_T$ in all centralities, similar to
inclusive hadrons and normal jet fragmentation measured in
$e^+e^-$ collisions. But the increase is a little bit stronger in
central Au+Au collisions as well as for the away side. The
$p/(\pi+k)$ ratio is about 0.1 for pure jet fragmentation and
$\sim 0.4$ for inclusive spectra in central Au+Au collisions. The
$p/(\pi+k)$ ratio for the away side jet is somewhere in between.
If the Mach-cone scenario is true, it is conceivable that the jet
excites the medium at the away side, which then fragments in the
usual recombination prescription. This would naturally lead to the
centrality dependent baryon/meson ratio in the away side.
\begin{figure}[h]
\epsfig{file=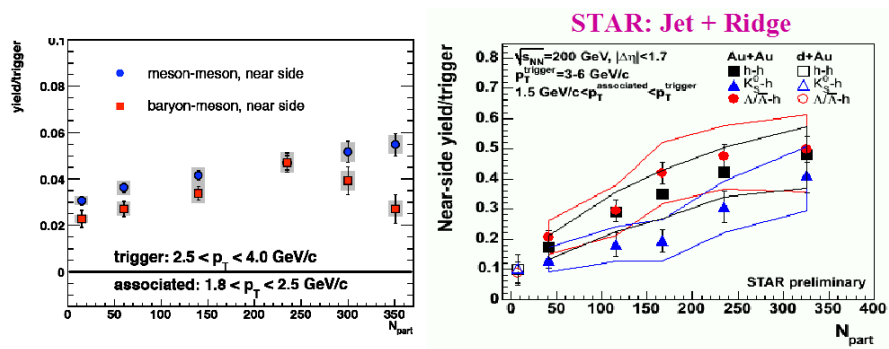,width=1.0\columnwidth}
\caption{\label{fig:id1} (left) the near PTY in $|\Delta\phi| <
0.94$rad for baryon-meson (squares) and meson-meson (circles)
correlations as a function of $N_{\rm{part}}$. (right) Similar
results from STAR.}
\end{figure}

Fig.\ref{fig:id1} shows the near side yield where the trigger is
identified in PHENIX~\cite{Adare:2006nn}(left) compared with
results from STAR (right)~\cite{Bielcikova:2007mb}. PHENIX Results
show a steady increase with centrality, except a quick drop of the
yield for trigger baryon at $N_{\rm{part}}>$250. The increase
trend is qualitatively consistent with results obtained from
hadron-hadron correlation (see Fig.\ref{fig:inte}), which was
argued to be due to the ridge in $\Delta\eta$. Similar trend is
also observed by STAR in the right panels, except that in their
case, the yield for trigger baryon is significantly larger than
for trigger mesons and yield for trigger baryon does not drop in
central collisions. However we should note that the trigger $p_T$
in STAR is different from PHENIX and they also include more ridge
contribution due to a much larger $\Delta\eta$ coverage. In any
case, a good theoretical model should be able to simultaneously
explain single particle yields, elliptic flow and as well as these
PID correlation results. Quark recombination (thermal-thermal and
thermal-shower) was quite successful in describing the spectra and
elliptic flow, but so far was not very successful in describing
the jet correlation yield, especially the centrality
dependence~\cite{Adare:2006nn}.

\subsection{Comments on $v_2$ background subtraction}
One of the important systematic errors in the jet correlation
comes from elliptic flow background subtraction (see
Eq.\ref{eq:core1}). In the past years, many different methods have
been developed to measure the $v_2$: Reaction-Plane (RP) Method
($v_2\{\rm{RP}\}$), 2 particle commulant method ($v_2\{2\}$), 4
particle commulant method ($v_2\{4\}$), and SMD-ZDC method
(Reaction-plane for direct flow $v_1$). These measurements have
quite different sensitivities on the non-flow effects and
eccentricity
fluctuation~\cite{Miller:2003kd,Zhu:2005qa,Manly:2005zy,Bhalerao:2006tp}.
The obvious question to ask is what $v_2$ one should use in the
jet correlation analysis?

The two particle correlation method and two particle commulant
method are closely related. Naturally, all non-flow effects and
event by event fluctuations that affects the commulant $v_2$ would
also contribute to the two particle azimuthal correlation. Thus it
seems that $v_2\{2\}$ should be used in the background subtraction
except that one would like to remove the jet bias to $v_2\{2\}$
since it is the signal. PHENIX use the $v_2\{\rm{RP}\}$ in all
correlation analysis. The $v_2\{\rm{RP}\}$ is determined by the
BBC which sits in the forward region ($3<|\eta|<4$). The RP method
measures separately the average $v_2$ for the triggers ($\langle
v_{2}^t\rangle$) and partners ($\langle v_{2}^a\rangle$). Due to
eccentricity fluctuation, the $v_2$ of the triggers and partners
fluctuate in the same direction event by event. This intrinsic
fluctuation would leads to $\langle
v_{2}^tv_{2}^a\rangle\neq\langle v_{2}^t\rangle\langle
v_{2}^a\rangle$. Both PHOBOS~\cite{Loizides:2007rm} and
STAR~\cite{Sorensen:2006nw} found a substantial eccentricity
fluctuation, around 40\%. From it, the additional correction
factor can be roughly estimated to be (assuming a gauss shape for
the fluctuation):
\begin{equation}
\label{eq:8} \sqrt {\left\langle {v_2^2 } \right\rangle }  \approx
\left\langle {v_2 } \right\rangle \left( {\sqrt {1 + \left(
{\frac{{\delta v_2 }} {{\left\langle {v_2 } \right\rangle }}}
\right)^2 } } \right) \approx 1.08\left\langle {v_2 }
\right\rangle
\end{equation}
So this suggest that the $v_2\{\rm{RP}\}$ should be increased by
8\% in order to account for the correlated fluctuation between
triggers and partners. However, PHENIX have measured both the
$v_2\{2\}$ $v_{2}\{\rm{RP}\}$~\cite{Adler:2004cj}. The agreements
between the two methods are better than 8\%. Note that
Eq.\ref{eq:8} is very sensitive to the $\delta v_2$ used in the
calculation and as well as the gauss smearing assumption. The
correction factor would be only 4.5\% for a 30\% input $v_2$
fluctuation. So probably the estimation of Eq.\ref{eq:8} is too
simplistic.

In many jet correlation analyses in STAR, the used $v_2$ is an
average between TPC $v_{2}\{\rm{RP}\}$ and $v_2\{4\}$. Since the
$v_2\{4\}$ reduces non-flow as well as fluctuation effects, it is
smaller than $v_2\{2\}$. This probably explains why their away
side jet shape at intermediate $p_T$ is less concave comparing to
PHENIX.
\begin{figure}[ht]
\begin{center}
\epsfig{file=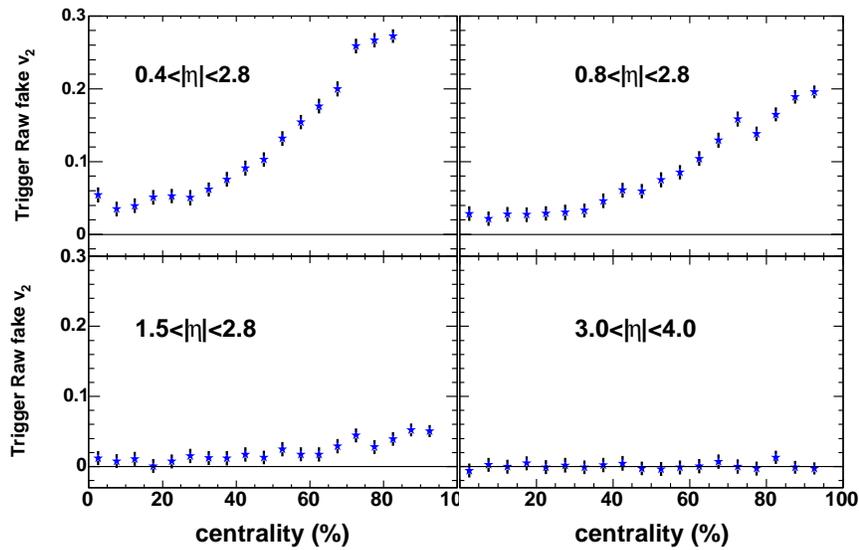,width=0.9\columnwidth}
\caption{\label{fig:v2bias} The fake $v_2$ of the leading particle
from jet as function of centrality using the EP determined in four
$\eta$ windows. The embedded jet is at mid-rapidity
($|\eta|<$0.35).}
\end{center}
\end{figure}

One of the advantages of the PHENIX RP $v_2$ measurement comes
from the large pseudo-rapidity gap between BBC and the central
arm. This gap greatly reduce the effects of the jet bias on the RP
$v_2$. PHENIX have perform detailed Monte Carlo simulation to
estimate the bias effect. Fig.\ref{fig:v2bias} shows the
centrality dependence of jet induced fake $v_2$ for various
pseudo-rapidity window used to determine the RP. The fake $v_2$
decreases as the $\eta$ window used to determine the RP angle is
further away from mid-rapidity. For the $\eta$ window in BBC
acceptance ($3<|\eta|<4$), the fake $v_2$ is almost negligible.
Further details can be found in~\cite{Jia:2006sb}.

\begin{figure}[ht]
\begin{center}
\epsfig{file=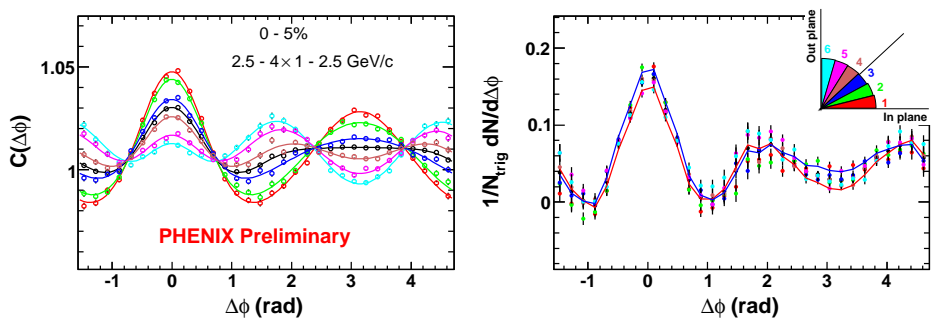,width=1.0\columnwidth}
\caption{\label{fig:rp} a) Correlation function for various 6
trigger direction bin and the trigger integrated bin (the center
black curve) in 0-5\% Au+Au at 200 GeV. b) The background
subtracted per-trigger yields, the insert figure shows the 6
trigger bins.}
\end{center}
\end{figure}

PHENIX have measured the jet yield as function of angle with
respect to the reaction plane at intermediate
$p_T$~\cite{Jia:2005ab}. Due to the smearing effect due to
relatively poor RP resolution of BBC and possibly a small jet
signal, we see little jet like difference of the per-trigger yield
in the in-plane direction relative to the out-plane direction(see
Fig.6 of~\cite{Jia:2005ab} and discussions.). We can use this
special property to constrain the $v_2$ simultaneously using CFs
measured at different angular direction. Fig.\ref{fig:rp}a shows
the CFs in 0-5\% central Au+Au collisions for 6 angular bins in
$15^o$ steps. Fig.\ref{fig:rp}b shows the jet yield after
subtracting the flow terms which can be calculated according to
simple equation~\cite{Jia:2005ab}. Although the CFs change
dramatically from in plane to out of plane, the calculated flow
term tracks the true flow background nicely. Given the small
eccentricity in 0-5\%, we can safely assume that the jet yield,
jet yield should not depend on the trigger direction. Thus the
relatively good agreements between 6 measurements is an
independent confirmation of the $v_2\{\rm{RP}\}$ used in the
correlation measurement. The small remaining difference between
the jet functions in Fig.\ref{fig:rp}b can be used to further
constrain the $v_2$.

\subsection{The meaning of per-trigger yield}
\label{sec:pty}
 In two particle correlation analysis, typically one
correlate particles in a high $p_T$ window (type $a$) with those
in a low $p_T$ window (type $b$). The distinction between triggers
and partners are arbitrary since medium modifications are on the
jet pairs. There is a simple relation between the PTY using the
high $p_T$ particles as triggers and that using low $p_T$
particles as triggers:
\begin{eqnarray}
\label{eq:eq1} R_{\rm{AA}}^aI_{\rm{AA}}^a   =
R_{\rm{AA}}^bI_{\rm{AA}}^b = \frac{{\rm{JetPairs}_{\rm{AA}}
}}{{N_{\rm{coll}} \times \rm{JetPairs}_{\rm{pp}} }} \equiv
J_{\rm{AA}}(p_T^a,p_T^b)
\end{eqnarray}
where JetPairs$_{\rm{AA}}$ and JetPairs$_{\rm{pp}}$ represent the
average number of jet pairs in one A+A collision and one p+p
collision, respectively. The $R_{\rm{AA}}^a$ represents
$R_{\rm{AA}}$ for type a particles and $I_{\rm{AA}}^a$ represents
PTY using type a particles as triggers. In naive jet quenching
picture, the observed triggers are biased close to the surface,
the corresponding away side companions have a larger medium to
traverse. One expects $I_{\rm{AA}}<R_{\rm{AA}}$ or
$J_{\rm{AA}}<R_{\rm{AA}}I_{\rm{AA}}$. In direct photon-jet
correlation, since direct photons are not modified by the medium,
the away side jet would behave exactly as the single jet
suppression: $I_{\rm{AA}} = R_{\rm{AA}}$.

In correlation analysis, we are interested in studying the
modification of the away side jet tagged by the triggering jet,
i.e the {\bf{per-jet yield}}. In reality we study the modification
on the {\bf{per-trigger yield}}, represented by $I_{\rm{AA}}$. The
physics interpretation of this quantity are complicated by the
modifications on the triggers. In general there can be two types
of modifications of the triggers:
\begin{enumerate}
\item Additional triggers not coming from jet: such as
thermal-thermal recombination, boost of soft particles due to
flow. They tend to dilution the PTY.

\item Pure medium modifications of the triggers: jet quenching,
enhancement due to energy dissipation such as radiative
gluons/Mach-cone/Cherekov. They either does not affect or increase
the PTY.
\end{enumerate}

\begin{figure}[t]
\epsfig{file=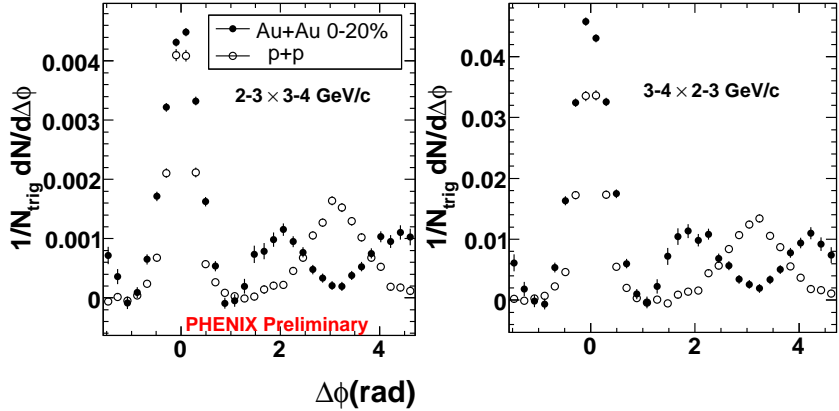,width=0.9\linewidth}
\caption{\label{fig:yi} The per-trigger yield for 2-3 and 3-4
GeV/$c$ correlation, a) using 2-3 GeV/$c$ particles as trigger, b)
using 3-4 GeV/$c$ particles as trigger.}
\end{figure}

(2) represents the effects we hope to study, which are complicated
by (1). Both can change $I_{\rm{AA}}$, but relation
Eq.\ref{eq:eq1} always holds. Fig.\ref{fig:yi} illustrates th idea
by showing the PTYs from Au+Au and p+p, where the two hadrons are
selected from non-overlapping $p_T$ ranges. In the left panel, 2-3
GeV/$c$ hadrons and 3-4 GeV/$c$ hadrons are used as triggers (a)
and partners (b), respectively. In the right panel, the role of
triggers and partners are swapped. Clearly, the vertical scale of
the left panel is less than that in the right panel, simply
because there are many more hadrons in 2-3 GeV/$c$ than 3-4
GeV/$c$. In the left panel, the nearside Au+Au data points are
close to those for p+p; whereas in the right panel, the nearside
Au+Au data are higher. This is the case because $R_{\rm{AA}}$ is
smaller in 3-4 GeV/$c$ than in 2-3 GeV/$c$. More suppression means
less number of triggers for fixed number of pairs, thus the PTY
increase relative to the p+p value.

Either (1) or (2) can explain this example. We can argue that most
of type-a particles come from recombination without jet
correlation, thus $I_{\rm{AA}}$ for type-a is smaller than that
for type-b. We can also argue that type-a particles come from jet
but are less suppressed than type-b, thus $I_{\rm{AA}}$ for type-b
is larger than that for type-a. $v_2$ and spectra measurements
suggest that a large fraction (more than 50\%) of the particles in
2-4 GeV/$c$ come from recombination. However the measured
per-trigger yield is significantly enhanced at $p_T<$4 GeV/$c$ as
shown by Fig.\ref{fig:inte}. This suggests that either triggers
are dominated by the shower-thermal recombination, hence
maintaining the jet correlation, or the enhancement of the jet
multiplicity is much bigger than the dilution to the triggers. PID
correlation measurement can help to clarify this issue.
Fig.\ref{fig:id1}~\cite{Adare:2006nn} indeed shows some
differences between baryon and meson triggers on the near side and
possibly also on the away side. This trend is not consistent with
the dilution from pure thermal-thermal recombination model
predictions~\cite{Adare:2006nn}. Further high statistics detailed
measurement of the PID correlation in intermediate $p_T$ can help
clarify the origin of the enhancement.

\section{Jet quenching}
\subsection{hadron-hadron and $\pi^0$-hadron correlation at high $p_T$}

\begin{figure}[th]
\epsfig{file=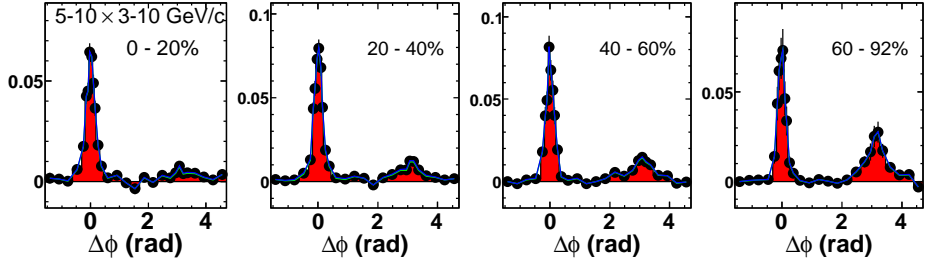,width=1.0\columnwidth}
\caption{\label{fig:scan} The hadron - hadron PTY as function of
centrality at high $p_T$ in 200 GeV Au+Au collisions.}
\end{figure}

At high $p_T$, the situation is a lot more cleaner. The dilution
to jet yield due to recombination or flow (if there is any) become
negligible. Since each jet fragments into one high $p_T$ trigger,
the per-trigger-yield at the away side is the same as the per-jet
yield. Fig.\ref{fig:scan} show the centrality dependence of the
jet yield for 5-10 GeV/$c$ trigger and 3-10 GeV/$c$ partner in 200
GeV Au+Au collisions. The shape and yield at the near side show
very little centrality dependence. The away side jet is clearly
peaked at $\pi$, but the magnitude is significantly suppressed in
central Au+Au collisions comparing to that in peripheral
collisions, consistent with strong quenching or absorption of the
away side jet.

PHENIX also carried out detailed study of the jet correlation
analysis in Cu+Cu collisions at 200 GeV. The jet correlation
measurements in Cu+Cu offer several unique advantages. In terms of
$N_{\rm{part}}$, Cu+Cu collisions cover from peripheral to 30-40\%
central Au+Au collisions. $N_{\rm{part}}$ can be determined with
better precision that in Au+Au, thus allows a more detailed
mapping of the centrality dependence of the onset of the jet
quenching. Secondly, systematic error from $v_2$ on the jet yield
is much smaller in Cu+Cu due to a smaller combinatoric background.

\begin{figure}[h]
\epsfig{file=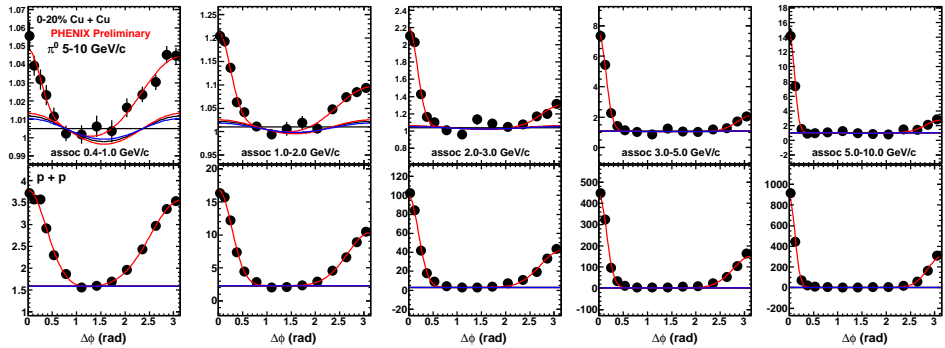,width=1.0\columnwidth}
\caption{\label{fig:cucucf} The correlation function in Au+Au
(upper panels) and p+p (bottom panels).}
\end{figure}

\begin{figure}[h]
\epsfig{file=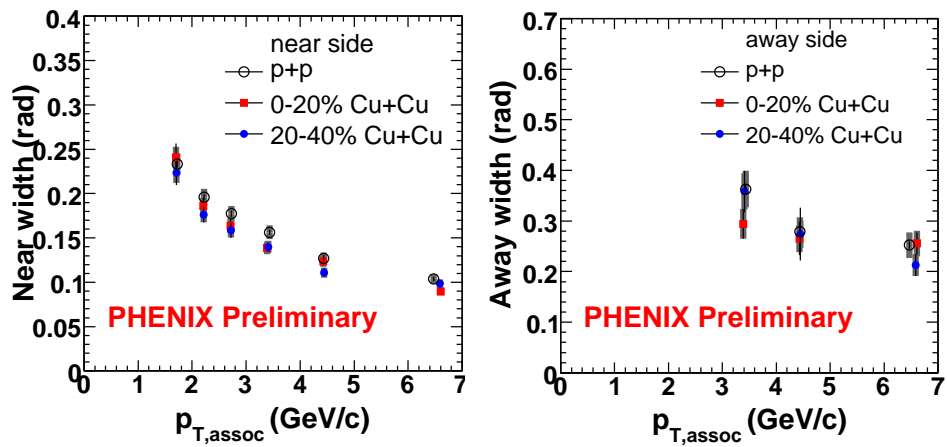,width=1.0\columnwidth}
\caption{\label{fig:cucuwidth} The near side and away side jet
gauss width for p+p, 0-20\% and 20-40\% Cu+Cu collisions.}
\end{figure}

The measurement was carried with 5-10 GeV/$c$ $\pi^0$ as triggers
and charged hadrons in several ranges in 0.4-10 GeV/$c$ as
partners. Fig.\ref{fig:cucucf} show the correlation function in
0-20\% Cu+Cu compared with that in p+p. A clear away side excess
can be seen in all partner $p_T$ ranges. No concave shape is
expected even after the $v_2$ background subtraction. At partner
$p_T>2$ GeV/$c$, the away side can be well fitted with a gauss
function. Fig.\ref{fig:cucuwidth} shows the near side and away
side jet width as function of partner $p_T$ for p+p, 0-20\% and
20-40\% Cu+Cu collisions. No apparent differences can be seen
between Cu+Cu and p+p, suggesting partners come mostly from jet
fragmentation in the $p_T$ range under consideration (all $p_T$ on
the near side and $p_T>$2 GeV/$c$ on the away side).

\begin{figure}[ht]
\epsfig{file=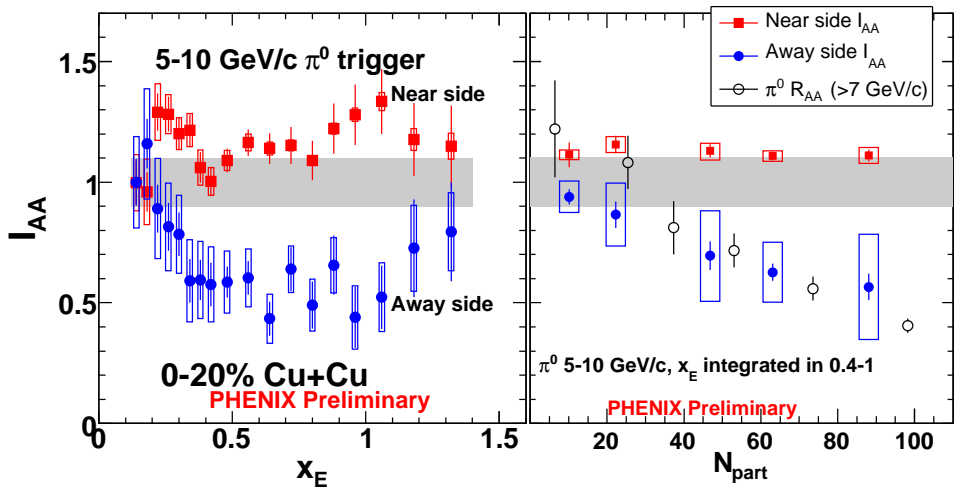,width=1\columnwidth}
\caption{\label{fig:xe}a) The $I_{\rm{AA}}$ as function of $x_E$
in p+p and Au+Au for near and away side; b) The $I_{\rm{AA}}$ for
yield integrated in 0.4-1 as function of $N_{part}$, compared with
suppression factor for high $p_T$ $\pi^0$.}
\end{figure}

Comparing the top and the bottom panels of Fig.\ref{fig:cucucf},
one notices that there is a larger asymmetry between the near and
away side in Cu+Cu than in p+p. To quantify the medium
modifications, we extract the PTY for Cu+Cu and p+p and construct
the $I_{\rm{AA}}$ as function of $x_E$ ($x_E =
p_{T,A}/p_{T,T}\cos(\Delta\phi)$). The results from 0-20\% central
Cu+Cu collisions are shown in the left panel of Fig.\ref{fig:xe}.
$I_{\rm{AA}}$ at the near side is consistent with 1 in the full
$x_E$ range of 0.1-1.4; the away side $I_{\rm{AA}}$ starts at
slightly above 1 at small $x_E$, and gradually decreases towards
larger $x_E$, at $x_E>0.4$ the suppression value approaches a
constant of 0.5. This constant behavior is also seen by STAR in
Au+Au collisions with $N_{\rm{part}}$~\cite{Adams:2006yt}. The
level of suppression was found to be similar to single particle
$R_{AA}$. Since the single particle $R_{AA}$ in Au+Au and Cu+Cu
collisions follows $N_{\rm{part}}$ scaling, we can ask whether
$I_{\rm{AA}}$ also has similar scaling behavior. The right panel
of Fig.\ref{fig:xe} shows the integrated $I_{\rm{AA}}$ in
$0.4<x_E<1$ as function $N_{\rm{part}}$, comparing with the
integrated $R_{AA}$ for the single particle. Assuming $\langle
z\rangle\sim0.7$ for leading $\pi^0$s, the original jets should be
around 5/0.7=7 GeV/$c$. Thus the $I_{\rm{AA}}$ of the away side
should be directly comparable to the single particle at $p_T>7$
GeV/$c$. In reality, since the $\pi^0$ $R_{\rm{AA}}$ is flat at
$p_T>4$ GeV/$c$, the exact values of the jet energy for high $p_T$
$pi^0$s are not important.
Near side $I_{\rm{AA}}$ is around one in all centralities,
consistent with surface emission picture. Away side $I_{\rm{AA}}$
shows a suppression that has a similar centrality dependence in
$N_{\rm{part}}$ relative to single particle suppression. This is
rather surprising given that the away side jet travels more medium
than the single particles in the naive jet absorption picture.
Probably suggesting that the simple geometrical bias argument in
Section\ref{sec:pty} is too naive. In jet quenching picture, the
suppression is due to the energy degradation of the high $p_T$
jets. The suppression factor $R_{\rm{AA}}$ depends on both the
energy loss itself as well as on the input parton spectra shape.
For the single particle, the expected $N_{\rm{binary}}$ scaled p+p
spectra have a typical power-law shape with a power of 8. In the
di-hadron correlation, the away side spectra associated with the
leading particles have much flatter distribution with a much
smaller power. Thus for the same amount of energy loss for single
jet and away side jet conditional to the near side jet, the
suppression level observed for $I_{\rm{AA}}$ could be well less
than the $R_{\rm{AA}}$. If we follow the prescription in
Ref.~\cite{Adler:2006bw}, the fractional energy loss
$S_{\rm{loss}}$ is related to $R_{\rm{AA}}$ as:
\begin{eqnarray}
S_{\rm{loss}} = 1 - R_{\rm{AA}}(p_T)^{1/(n-1)} \quad or \quad
R_{\rm{AA}} = (1-S_{\rm{loss}})^{n-1}
\end{eqnarray}
The power ``n'' is $7.1$ for single particle spectra in $dN/dp_T$.
The power for the away side associated spectra can be determined
from $p+p$ data~\cite{Adler:2005ad} and it is about 4.8 in
$dN/dp_T$. If the away side jet have same fraction energy loss as
single particle, then we would expect:
\begin{eqnarray}
S_{\rm{loss}} = 1 - R_{\rm{AA}}(p_T)^{1/(n_{R}-1)} =1 -
I_{\rm{AA}}(p_T)^{1/(n_{I}-1)}.
\end{eqnarray}
$R_{\rm{AA}} =0.2$ in central Au+Au collisions would lead to
$I_{\rm{AA}} = R_{\rm{AA}}^{(n_I-1)/(n_{R}-1)}= 0.37$, much bigger
than $R_{\rm{AA}}$. On the other hand, if we require $I_{\rm{AA}}=
R_{\rm{AA}}$, then the away hadron energy loss fraction would be
$S^{I}_{\rm{loss}} = 1 - I_{\rm{AA}}(p_T)^{1/(n_{I}-1)} = 0.345$,
much bigger than the single hadron energy loss fraction
$S^{R}_{\rm{loss}}=0.23$, as expected (about 50\% more energy
loss). If we apply the same trick to central Cu+Cu collisions,
assuming $I_{\rm{AA}} = R_{\rm{AA}} =0.5$, the fractional energy
loss is would be $S^{R}_{\rm{loss}} = 0.107$ and
$S^{I}_{\rm{loss}} = 0.167$ for single particle and away side jet,
respectively.

\subsection{$\gamma$-hadron correlation}

In high $p_T$ hadron - hadron or $\pi^0$-hadron correlation, the
total jet energy is not known. Trigger hadron from normal jet
fragmentation carries on average about 60-70\% ($\langle
z\rangle$) of the total jet energy. This effect is known as the
trigger bias, which have been studied in detail in p+p and d+Au
collisions~\cite{Adler:2006sc,Adler:2005ad}. The energy loss of
triggering jet in the medium leads to additional bias effects, and
since the energy loss is dependent on path length, the trigger
bias effects would also coupled with the collision geometry. Thus
complicates the extraction of the medium effects. Direct photon
($\gamma_{\rm{dir}}$)- hadron correlation does not suffer from
such limitations. $\gamma_{\rm{dir}}$ serves as the gauge of the
jet energy and jet direction, allowing the direct study of the
medium modification of the away side jets. In reality,
$\gamma_{\rm{dir}}-h$ correlation has its own difficulties and
limitations. In particular, once have to deal with large decay
$\gamma$ background, non-hard-scattering contributions such as the
fragmentation and bremsstrahlung and thermal radiation and small
rate.

\begin{figure}[th]
\begin{center}
\epsfig{file=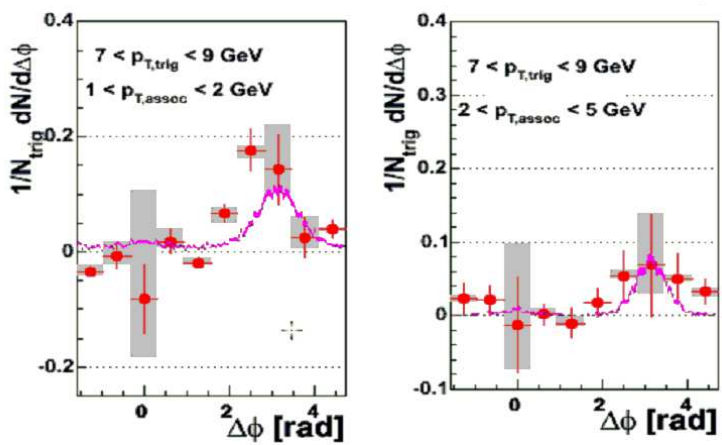,width=0.8\columnwidth}
\caption{\label{fig:gammapp} The direct $\gamma$-h signal in p+p
collisions compared with PYTHIA simulation for 7-9 GeV/$c$ trigger
$\gamma$ and two partner $p_T$ ranges.}
\end{center}
\end{figure}

PHENIX has observed a $p_T$ dependent direct photon excess above
the decay background in central Au+Au collisions. A statistical
subtraction method combined with the knowledge on the direct
$\gamma$ excess~\cite{Adler:2006yt} is used to obtain the
$\gamma_{dir}-h$ signal. Fig.\ref{fig:gammapp} show the hadron
yield associated with direct photons in $p+p$ collisions for two
partner $p_T$ ranges. Near side is consistent with zero, and away
side shows some finite excess. The p+p data are compared with the
expected values from the PYTHIA6.1 calculation with $k_T$=2.5
GeV/$c$. Within the large systematic uncertainties, the data
qualitatively agrees with the PYTHIA value.

\begin{figure}
\epsfig{file=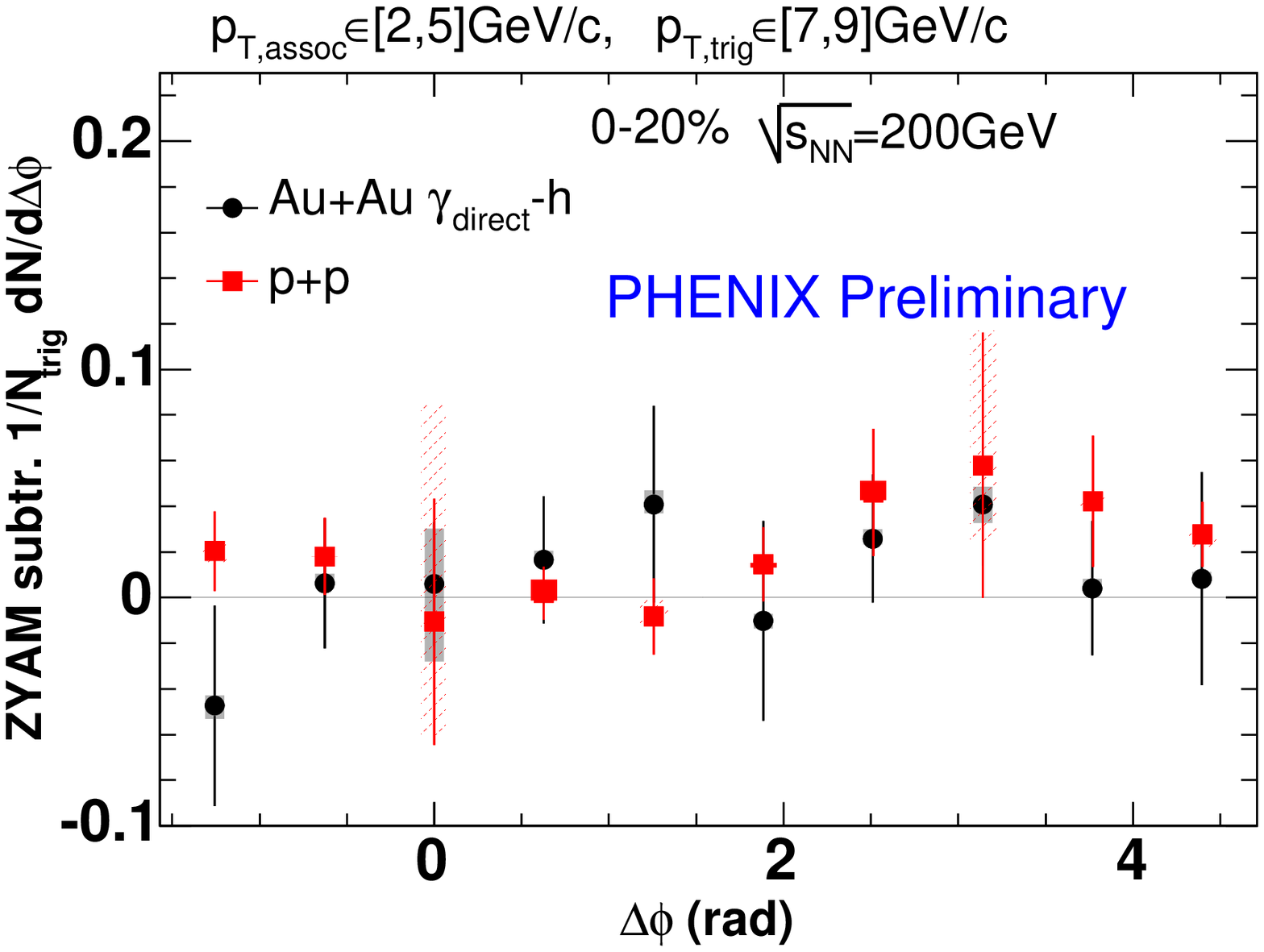,width=0.5\columnwidth}\epsfig{file=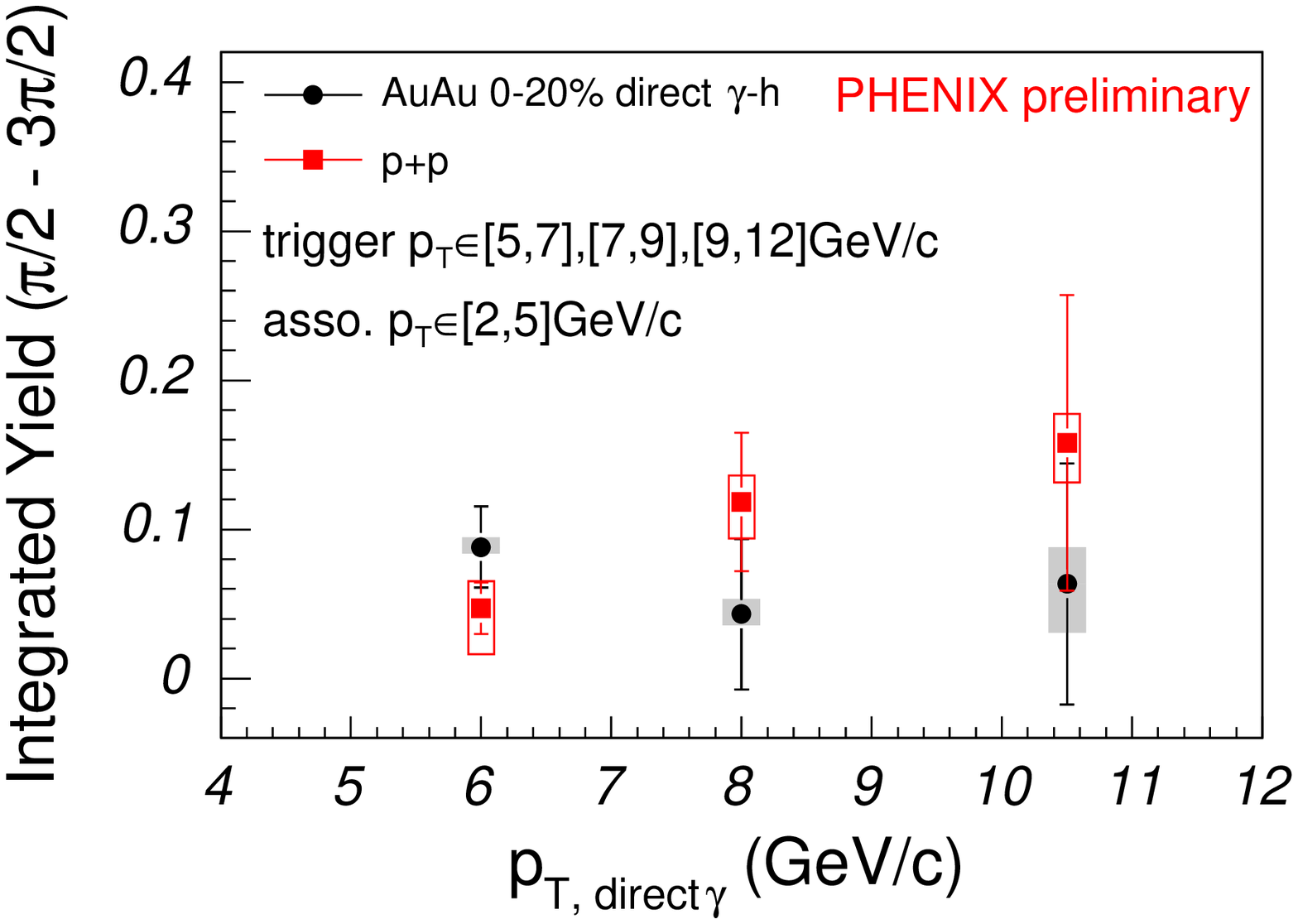,width=0.5\columnwidth}
\caption{\label{fig:gammaauau}a) The per-trigger yield for direct
$\gamma$ in central Au+Au (circles) and p+p (boxes). b) The
integrated yield of the away side for several trigger direct
$\gamma$ ranges (data are plotted in the middle of the bin)}
\end{figure}

Left panel of Fig.\ref{fig:gammaauau}~\cite{Jiamin} shows the
comparison of the $\Delta\phi$ distribution of the PTY in p+p and
0-20\% central Au+Au collisions. The Au+Au data have smaller
systematic errors, however the measurements are currently
statistics limited, especially at around $\Delta\phi=\pi/2$
region. The Au+Au data seems to show a small positive yield at
$\pi$, but the significance is only about 1 $\sigma$ level. One
can reduce the statistic errors by integrating the away side yield
in $\pi/2<\Delta\phi<3\pi/2$. The right panel of
Fig.\ref{fig:gammaauau} show the integrated yield for several
trigger $p_T$ regions with partner $p_T$ fixed in 2-5 GeV/$c$. p+p
data suggest a gradual increase of the PTY towards larger trigger
$p_T$, while the Au+Au data seems to stay relative flat and below
the p+p level.

Overall, the results of the direct $\gamma-h$ correlation is still
in a very qualitative stage, mainly due to limited statistics.
However, the necessary methodologies have been developed and
tested by PHENIX. It is only a question whether there will be
sufficient statistics from future Au+Au runs. We might need the
RHIC-II in order to do a really jet tomography measurement with
direct photons.

\section{Conclusion and outlook}

PHENIX have mapped out the landscape of two particle azimuth
correlation in heavy ion collisions as function of $p_{T,T}$,
$p_{T,A}$, centrality, hadron species, $\sqrt{s}$ and collision
systems in different $\Delta\phi$ regions. The background
subtracted ``jet'' pairs seems to come from four different
components concentrated in different $\Delta\phi$ regions with
their characteristic dependence on $p_T$, PID, $\sqrt{s}$ etc. 1)
A hard component around $\Delta\phi=0$ that is consistent with jet
fragmentation; 2) A soft and broad component around $\Delta\phi=0$
that is consistent with the ``Ridge'' seen by STAR; 3) A hard
component (``Head'') around $\Delta\phi=\pi$ consistent with
fragmentation of away side jet; and 4) a soft component centered
at $|\Delta\phi-\pi|\sim1$ (``Shoulder''). The hard components are
sensitive to the ``quenching'' of the (di-)jets by the medium. The
hard component at the near side shows little modification, whereas
the hard component at the away side are strongly suppressed, up to
factor of 5 in central Au+Au collisions. The soft components
(``Ridge'' and ``Shoulder'') represent the response of the medium
to the jets, appearing as distortions of the shape and
enhancements of the yield. These components are shown to be
important at $p_T<4$ GeV/$c$ and have a particle composition that
is closer to the bulk medium. The non-trivial evolutions of the
near and away side shape/yield with $p_T$ and $\sqrt{s}$ are the
results of the detailed interplay between the soft and hard
components.


The four different components can have very different geometrical
bias and trigger bias. In a simple jet absorption picture where a
very opaque medium is assumed, the near side hard component is
emitted from the surface, $\langle L\rangle\sim 0$; the near side
``Ridge'' comes from jet close to the surface, $\langle
L\rangle<R$ ($R$ is the average radius of the medium); the away
side ``Shoulder'' results from jet that traverses a large path
length, $\langle L\rangle>R$; the path length for away side hard
component can vary dramatically depends whether it is tangential
emission or punch-through, $0<\langle L\rangle<2R$. One of the
important future directions of the jet correlation is to
quantitatively separate the four different components; and to
study the detailed properties of each components. To fully
understand the medium response, one need to address the question
that whether the near side ``Ridge'' and away side ``Shoulder''
are of the same origin or not. A detailed mapping of the $p_T$,
PID, charge and $\sqrt{s}$ dependence would be very helpful in
this regard. One can dial the path length by triggering on two
high $p_T$ triggers and correlate with the third soft hadrons, the
``displaced'' peak might be seen on both the near and away
side~\cite{Renk:2006pk}. However this is only possible if the high
$p_T$ have a significant punch-through component and the medium
can not be very opaque. On the jet quenching part, high $p_T$
correlation remains to be a good tomographic tool since it is less
affected by the surface bias~\cite{Zhang:2007ja}. Both the
centrality and reaction plane dependence of the jet correlation at
high $p_T$ would be very helpful in constraining the transport
properties of the medium. In the meanwhile, we should continue to
pursue the gamma-jet correlation with increased statistics and
refined analysis techniques. Since it is less affected by the
geometrical bias and trigger bias, even a statistics limited
result could be very powerful in constraining various jet
quenching models~\cite{Renk:2006qg}.

%
%
%
%

\end{document}